\documentclass[prb,twocolumn,showpacs,showkeys,superscriptaddress]{revtex4}
\usepackage{graphicx}
\usepackage{dcolumn}
\usepackage{eucal}
\usepackage[dvips]{epsfig}
\usepackage{changebar}
\usepackage{amssymb}
\usepackage{amsfonts}
\usepackage{amsmath}

\begin{document}

\title{Ferromagnetism in $Co$-doped $ZnO$ films grown by molecular beam epitaxy: magnetic, electrical and microstructural studies}

\author{V.V.~Strelchuk}
\address{V. Lashkaryov Institute of Semiconductor Physics National Academy of Sciences of Ukraine,
45 Nauky pr., 03028 Kyiv, Ukraine}
\author{V.P.~Bryksa}
\address{V. Lashkaryov Institute of Semiconductor Physics National Academy of Sciences of Ukraine,
45 Nauky pr., 03028 Kyiv, Ukraine}
\author{K.A.~Avramenko}
\address{V. Lashkaryov Institute of Semiconductor Physics National Academy of Sciences of Ukraine,
45 Nauky pr., 03028 Kyiv, Ukraine}
\author{P.M.~Lytvyn}
\address{V. Lashkaryov Institute of Semiconductor Physics National Academy of Sciences of Ukraine,
45 Nauky pr., 03028 Kyiv, Ukraine}
\author{M.Ya.~Valakh}
\address{V. Lashkaryov Institute of Semiconductor Physics National Academy of Sciences of Ukraine,
45 Nauky pr., 03028 Kyiv, Ukraine}

\author{V.O.~Pashchenko}
\address{B. Verkin Institute for Low Temperature Physics and Engineering National Academy of Sciences of Ukraine, 47 Lenin Ave., 61103 Kharkiv, Ukraine}
\author{O.M.~Bludov}
\address{B. Verkin Institute for Low Temperature Physics and Engineering National Academy of Sciences of Ukraine, 47 Lenin Ave., 61103 Kharkiv, Ukraine}

\author{C.~Deparis}
\address{Centre de Recherches sur l'H\'{e}t\'{e}ro\'{e}pitaxie et ses Applications, CNRS, F-06560 Valbonne, France}
\author{C.~Morhain}
\address{Centre de Recherches sur l'H\'{e}t\'{e}ro\'{e}pitaxie et ses Applications, CNRS, F-06560 Valbonne, France}

\author{P.~Tronc}
\address{Centre National de la Recherche Scientifique, Ecole Superieure de Physique et de Chimie Industrielles de la Ville de Paris, 10 rue Vauquelin, 75005 Paris, France}
\date{\today}

\begin{abstract}
We studied structural, optical and magnetic properties of high-quality $5$ and $15\%$ $Co$-doped $ZnO$ films grown by plasma-assisted molecular beam epitaxy (MBE) on $(0001)$-sapphire substrates. Magnetic force microscopy (MFM) and magnetic measurements with SQUID magnetometer show clear ferromagnetic behavior of the films up to room temperature whereas they are antiferromagnetic below $200$ $K$ approximately. Temperature dependence of the carrier mobility was determined using Raman line shape analysis of the longitudinal-optical-phonon-plasmon coupled modes.
It shows that the microscopic mechanism for ferromagnetic ordering is coupling mediated by free electrons between spins of Co atoms. These results bring insight into a subtle interplay between charge carriers and magnetism in MBE-grown $Zn_{1-x}Co_xO$ films.
\end{abstract}

\maketitle

\section{Introduction}

Currently one can observe a great interest in understanding and designing the physical properties of diluted-magnetic-semiconductor (DMS) structures. Indeed, they have potential applications in spintronics, where controlling electron spin can give rise to new devices. Since theoretical calculations predicted possible room-temperature ferromagnetism (FM)~\cite{Dietl}, in transition-metal-doped $Zn_{1-x}T_xO$ films ($T=Cr^{2+}$, $Mn^{2+}$, $Fe^{2+}$, $Co^{2+}$ and $Ni^{2+}$) they attracted a great interest. Experimental observations of room-temperature FM in $V^{2+}$:ZnO~\cite{Saeki}, $Fe^{2+}$:ZnO~\cite{Han}, and $Co^{2+}$:ZnO~\cite{Ueda} appeared in the literature. It seems reasonable to assume that FM is merely due to magnetic impurities, even if some experimental results appeared to rule this out~\cite{Potzger}. At the present time, the microscopic mechanism responsible for high-$T_c$ FM is still quite controversial for the II-VI compounds, especially for $ZnO$ based DMSs~\cite{Snure}. Various mechanisms have been proposed for the bulk materials, for example, carrier-induced ferromagnetism~\cite{Dietl} and percolation of bound magnetic polarons~\cite{Kaminski}. In addition,  structural defects probably play significant role in controlling the ferromagnetic properties of the $ZnO$. The main reasons referred to in the literature for appearance of a ferromagnetic phase are substitution of $Zn$ atoms by $Co$ ones and existence of magnetic clusters of metallic $Co$ and/or $Co$ oxides in $ZnO$ host. Furthermore, the magnetic properties of $Co^{2+}$:$ZnO$ films have a strong dependence on synthesis and processing conditions~\cite{Chambers}. In some cases, even the conclusion of intrinsic ferromagnetism remains controversial~\cite{Chambers}.

The ferromagnetic properties of $3d$-metal-doped $ZnO$ nano-particles were explained using the core-shell model~\cite{Wang}. High stability of ferromagnetic phase in $Ni^{2+}$:$ZnO$ nanocrystals was related to high surface-defect-concentration~\cite{Zhang}. High-$T_c$ ferromagnetism in $Mn^{2+}$:$ZnO$ and $Co^{2+}$:$ZnO$ nanocrystals was interpreted as a result of long-range exchange interaction of $Mn^{2+}$ and $Co^{2+}$ ions mediated by charge carriers~\cite{Kittilstved}. The important role of magnetic anisotropy of $Co^{2+}$ ions in $ZnO$ lattice~\cite{Sati} has been discussed as well as clearly observed correlation between magnetism and carrier concentration in $Zn_{1-x}Co_xO$ films~\cite{Kittilstved1}. Nano-scale non-uniform distribution of magnetic ions in the host lattice and spinodal decomposition have recently been observed in $Cr$-doped $ZnSe$ films ~\cite{Kuroda}. The films show ferromagnetic ordering with high values of the Curie temperature.

Three models have been proposed to explain room temperature ferromagnetism in $Zn_{1-x}Co_xO$ alloys. In the first one, ferromagnetism is mediated indirectly via free carriers (Ruderman-Kittel-Kasuya-Yoshida (RKKY) or double exchange mechanism model).  In the second one, ferromagnetism originates from secondary phase such as metallic $Co$ or $Co$-oxides. And the latter is due to the bound magnetic polaron model. In order to clarify this ambiguous situation, we studied MBE-grown $Zn_{1-x}Co_xO$ thin films with the help of  magnetic force microscopy (MFM), confocal micro-Raman, photoluminescence (PL) and $SQUID$ techniques.

Raman scattering became a very useful and informative technique for studying different phonon excitations in undoped and doped by $Li$, $N$, $Fe$, $Sb$, $Ga$, $Al$ $ZnO$ films. It allows studying influence of structural disorder in $ZnO$ lattice on vibrational properties~\cite{Bundesmann}. The study of $Co-O-Zn$ local vibration modes versus concentration of oxygen vacancies~\cite{Sudakar} allows correlating  carrier concentration and magnetic properties. Appearance of phonon bands at $186$, $491$, $526$, $628$, and $718$ $cm^{-1}$ was interpreted as a signature  of $Zn_yCo_{3-y}O_4$~\cite{Samanta} spinel phase. Note, that nanometer-size $Zn_yCo_{3-y}O_4$ clusters can be very easily detected in micro-Raman measurements whereas X-ray diffraction method is not well-suited for studying so small clusters.

In the absence of magnetic secondary phases, the distribution of $Co^{2+}$ ions over cation sites of $ZnO$ lattice should play an important role for ferromagnetism. A substituting  $Co^{2+}$ ion at a $Zn$ site can have no Co second first neighbor i.e., not to be involved in one (several) Co-O-Co sequence(s)  or have at least one Co second neighbor. The magnetic properties have strong dependence on the number of Co atoms of the first type (isolated atoms). Assuming that Co atoms are randomly distributed over cation sites and neglecting antisite and interstitial-site occupation, it was shown for the $5$ and $15\%Co$-doped $ZnO$ films that $94$ and $14\%$ of $Co$ atoms, respectively, belong to the first type. Note that the assumption rules out metallic Co clusters~\cite{Chambers,Ney1}. It is expected that isolated $Co$ atoms have ferromagnetic interaction mediated by free carriers.  In the $Co-O-Co$ bonding configurations, the two neighboring $Co$ localized spins are assumed to be coupled antiparallel providing antiferromagnetic properties , especially at low temperature~\cite{Chambers,Ney1,Morhain,Ney,Kobayashi}. As a result there is a competition between the ferromagnetic and antiferromagnetic interactions in the $Zn_{1-x}Co_xO$ films. It is expected that  ferromagnetic interactions between $Co^{2+}$ ions should take place in high quality MBE-grown $Zn_{1-x}Co_xO$ films with high electron concentrations ($n>10^{19}$ $cm^{-3}$)~\cite{Kobayashi}.

The present paper is organized as follows. Section II describes the growth procedure and setup for  micro-Raman, MFM and magnetic measurements. In Section III, we focus on the magnetic, structural, optical and electronic properties of the MBE-grown $5\%$ and $15\%$ $Co$-doped $Zn_{1-x}Co_xO$ thin films. Using the MFM and SQUID technique, the magnetic interactions in the films are determined. 
In the films a broad emission peak at $1.816$ eV ($683$ nm) is put into evidence which is ascribed to electron transitions within substitional $Co^{2+}$ ions. These results confirm that the $Co^{2+}$ ions are located at the $Zn$ sites in the wurtzite $ZnO$ structure. The micro-Raman measurement confirms the crystalline wurtzite structure in $Co$-doped $ZnO$ films.The temperature-dependent Raman measurements of longitudinal-optical-phonon-plasmon coupled modes (LOPCM’s) are also provided. Modeling Raman spectra for the LOPCM’s allows determining the temperature dependence of the carrier mobility. The results evidence that the ferromagnetism of the $Zn_{1-x}Co_xO$ films is due to the free carriers with high mobility and supports an indirect interaction of localized magnetic moments of isolated $Co^{2+}$ ions in the $ZnO$ lattice. Section IV is assigned to summary and outlook.

\section{Experimental details}

The $Zn_{1-x}Co_xO$ films were grown on $c$-sapphire substrates in a Riber Epineat MBE system equipment with conventional effusion cells for elemental $Zn$ and $Co$. Atomic oxygen was supplied via an Addon radio-frequency plasma cell equipped with a high-purity quartz cavity~\cite{Morhain}. The film thickness is about $1.7$ $\mu m$. The epilayer crystalline quality is attested by low full-width-at-half-maximum values in high-resolution X-ray diffraction scans for high-symmetry as well as oblique directions (see Table~\ref{GParam}). Lattice parameters of the pure $ZnO$ sample matches well with the values of $ZnO$ single crystal ($a=3.2495$ $\AA{}$, $c=5.2069$ $\AA{}$). After $Co$ substitution with $Zn$ atom both $a$- and $c$-axis lattice constants are changed ($a=3.266$ $(3.259)$ $\AA{}$ and $c=5.197$ $(5.195)$ $\AA{}$ for $5$ ($15$) at.$\%$ $Co$). Some discrepancy between the concentrations determined using contactless sub-micrometer Raman and macro-Hall measurements (Table~\ref{GParam}) can be caused by differences in local regions of measurements and possible changes of electric parameters due to heating under contact formation for $Co$-doped $ZnO$ films.

\begin{table}[h]
\caption{\label{GParam}
Parameters of MBE growth for undoped and $Co$-doped of the $ZnO$ films and determined values of the carrier mobility and concentrations obtained by Hall and Raman measurements at room temperature.}
\begin{tabular}{ccccccc}
\hline
\hline
\\
Sample number & & 226 & & 283 & & 288 \\
\hline
$Co$ concentration  & &  & &  & & \\
 ($\%$) & & $-$ & & $5$ & & $15$\\
\hline
Growth rate  & &  & &  & & \\
 ($\mu m$/hour) & & $0.66$ & & $0.43$ & & $0.43$\\
\hline
Growth temperature  & &  & &  & & \\
 ($^\circ$C) & & $510$ & & $560$ & & $560$\\
\hline
X-ray line  & & $-$ & & [$002$]: $0.29$ & & [$002$]: $0.32$\\
 FWHM (degree) & &  & & [-$105$]: $0.28$ & & [-$105$]: $0.21$\\
 & &  & & [$102$]: $0.78$ & & [$102$]: $0.57$\\
& &  & & Twist: $\pm0.54$ & & Twist: $\pm0.35$\\
\hline
Mobility  & &  & &  & & \\
 ($cm^2/Vs$) & &  & &  & & \\
measured by & &  & &  & & \\
Hall  & & $32$ & & $47$ & & $29$\\
Raman*  & & $-$ & & $98$ & & $130$\\
\hline
Electron density  & &  & &  & & \\
 ($cm^{-3}$) & &  & &  & & \\
measured by & &  & &  & & \\
Hall  & & $1\times10^{18}$ & & $0.1\times10^{20}$ & & $0.7\times10^{20}$\\
Raman*  & & $-$ & & $1.2\times10^{20}$ & & $1.3\times10^{20}$\\
\hline
\hline
\end{tabular}
* The accuracy of determining the carrier concentration from analysis of modelled $\omega^+$ LOPCM was about $\pm20\%$.
\end{table}

Confocal micro-Raman and PL measurements were performed using the $488.0$  nm line of the $Ar^+/Kr^+$ laser and recorded with a Jobin-Yvon $T64000$ spectrometer equipped with a CCD detector. Spatial resolution (lateral and axial) was about $1$ $\mu m$. The temperature-dependent micro-Raman spectra ($80$-$500$ K) were performed using a Linkam THM$600$ temperature stage.

The MFM measurements were performed by a Dimension $3000$ Nano-Scope $IIIa$ scanning probe microscope for spatial mapping of the magnetization structure of the out-of-plane component of the magnetic stray field of the $Zn_{1-x}Co_xO$ sample surface at room temperature. Before measurements, the probe was magnetized using a strong permanent magnet with the field aligned along the tip axial axis. Then the MFM images of the sapphire substrate surface were taken and no magnetic signal was registered.
The magnetic force gradients were measured using a two-pass technique (Lift Mode), where the topography was scanned at the first pass in the tapping mode and then the magnetic field gradients were obtained using oscillation frequency shift of the probe moving over surface. The cobalt coated Veeco magnetic tips with a coercivity of $\thickapprox400$ Oe, magnetic moment of $1\times 10^{-13}$ $emu$ and $25$ nm nominal tip apex radius were used. The two opposite orientations of probe magnetization were used (i.e. North or South pole on the tip apex). This allows distinguishing the signal of the gradient magnetic fields from other artefacts of long-range electrostatic fields detected by the magnetic tip apex. The value of lift scan height was optimized for maximal sensitivity and minimal topography effects and was about $100$ nm. The chip structure of $Zn_{1-x}Co_xO$ samples were also studied by using a ZEISS EVO-$50$ scanning electron microscope (SEM). The magnetic measurements were carried out using a Quantum Design SQUID magnetometer $MPMS$-$XL5$.

\section{Results and Discussion}

\begin{figure}
\centerline{\includegraphics[angle=0,width=0.4\textwidth]{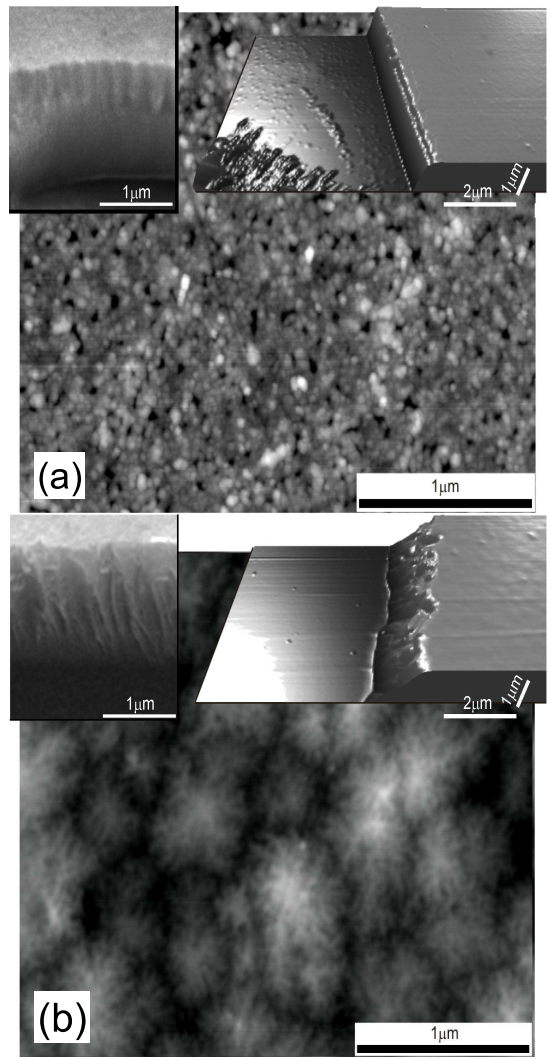}}
\bigskip
\caption{\label{fig01afm} $3\times3$ $\mu m^2$ AFM images for $ZnO$ films doped with $5\%$(a) and $15\%$(b) on the sapphire substrate. On insert show  $10\times10$ $\mu m^2$ SEM (left) AFM (right) images of the chipped edge.}
\end{figure}

AFM is used to characterize surface morphology, root-mean-square (rms) roughness, to verify the microstructures of $Zn_{1-x}Co_xO$ films. AFM morphology image of $Zn_{1-x}Co_xO$  films in the regions of chipped film edge changes with increase of the $Co$ concentration (Fig.~\ref{fig01afm}). As seen from Fig.~\ref{fig01afm}(a), morphology of $5\%$-doped $ZnO$ films is very tiny crystal grains ($20-30$ nm) due to the vertical columnar growth mode, and it shows the rms roughness of about $1.8$ nm. For $15\%$ $Co$-doped $ZnO$ films (Fig.~\ref{fig01afm}(b)) the rms roughness is close to $2$ nm, the larger domain structures with sizes from $100$ to $400$ nm are formed by connecting smaller crystal grains. Let us note that in this case the smaller crystal grain size is practically unchanged with increasing $Co$ concentration. Similar morphology was reported for the $Zn_{1-x}Co_xO$~\cite{Liu} and $Al$-doped $ZnO$~\cite{Dong} films.

\begin{figure}
\centerline{\includegraphics[angle=0,width=0.45\textwidth]{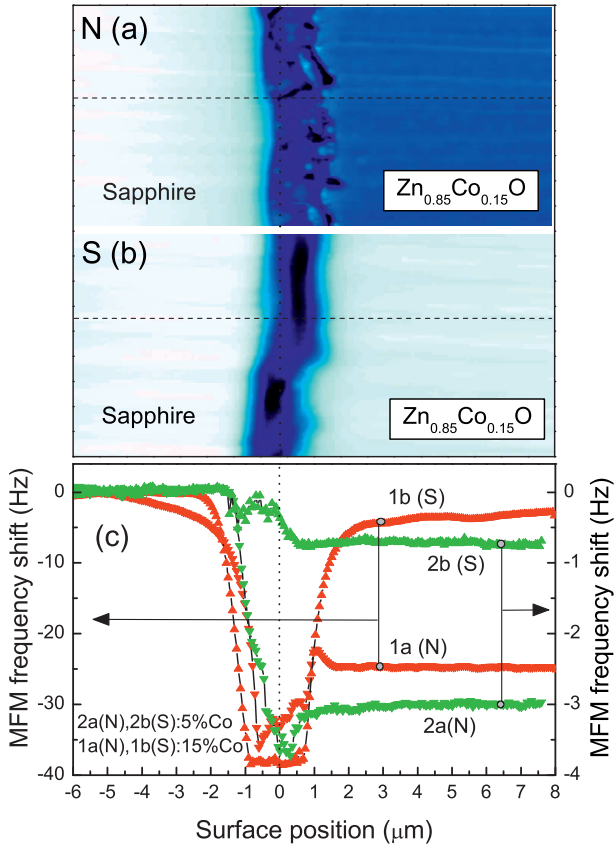}}
\bigskip
\caption{\label{fig01}MFM images at room temperature of the chipped edge of $15\%Co$-doped $ZnO$ films scanned under North (a) and South (b) tip apex magnetization (deep blue color corresponds to higher MFM signal). Profiles of surface magnetic field gradients (c) along horizontal dashed lines on the $(a)$ and $(b)$ magnetic maps recorded for the North(N) and South(S) tip apex magnetization with the $15\%$ (1a(N) and 1b(S) curves) and $5\%$ (2a(N) and 2b(S) curves) $Co$-doped $ZnO$ films.}
\end{figure}

The surface sensitive MFM method was used to study magnetization in the vicinity of the chipped sample edge of the $Co$-doped $ZnO$ films. We deal with the area of pure substrate and sharp film edge (Fig.~\ref{fig01}). As seen from the profiles (Fig.~\ref{fig01}c), magnetization of the $15\%Co$-doped $ZnO$ film, exhibits a sharp jump in magnetic signal at the chipped edge for the South and North pole of the probe (Fig.~\ref{fig01}a,b). The MFM magnetization map is independent of the AFM topography image of the $Zn_{1-x}Co_xO$ surface films. For the $5\%Co$-doped $ZnO$ film, similar changes in the MFM image take place, but changes are not so sharp, and their value is $\thicksim10$ times lower than for the $15\%Co$-doped $ZnO$ film(Fig.~\ref{fig01}c).

In the area of the film sharp edge, MFM probe interacts with the studied surface not only by the tip apex but by some area of side surface also, which causes local increase of the MFM signal amplitude. As shown on the Fig.~\ref{fig01}c, magnetization of the $Zn_{1-x}Co_xO$ films for the South pole has a lower value when compared with that of the North pole. This fact can be explained by the hysteresis of the magnetic film (curves 1b(S) and 1a(N) on Fig.~\ref{fig01}c) and agrees well with SQUID data at $T=300$ K and $H\parallel c$ geometry in a magnetic field of $H_{MFM}\approx 15 Oe$ (insert of the Fig.~\ref{fig02a}). It does indicate ferromagnetic behavior at room temperature in the films.

The MFM investigations did not reveal any fine magnetic surface structure of the films even in the high resolution mode with a lift height of $10$ nm. The observed uniform contrast of the MFM picture can testify for homogeneous distribution of the doping $Co$ impurity over the surface of the $Zn_{1-x}Co_xO$ alloys at least with the precision of our MFM experiments ($\>20$ nm). In the opposite case, the magnetic contrast and surface topography image would be correlated in some manner. It was observed, for instance, for $V^{2+}$:$ZnO$ nano-rods, when the pattern of separate vertically oriented magnetic dipoles were correlated with topographical AFM images of the nano-rods~\cite{Schlenkera}.

\begin{figure}
\centerline{\includegraphics[angle=0,width=0.45\textwidth]{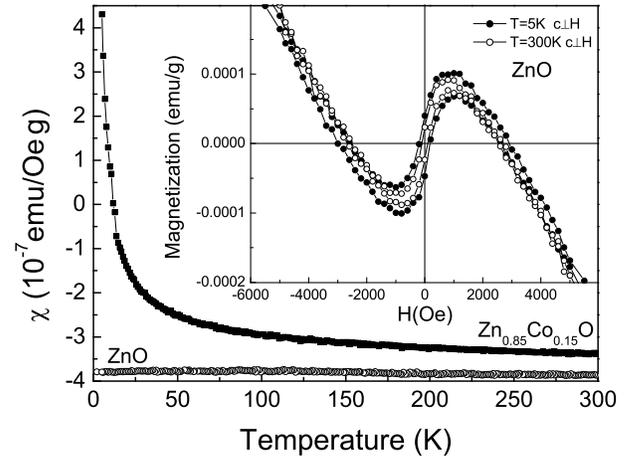}}
\bigskip
\caption{\label{fig02aa}Magnetic susceptibility of the $15\%Co$-doped $ZnO$ film (filled square) and undoped $ZnO$ film (open circle) versus temperature at a magnetic field of $H=1000$ Oe with $H\perp c$ geometry. The insert shows a dependence of the magnetization for the undoped $ZnO$ film versus magnetic field at $5$ (filled circles) and $300$ K (open circles) for $H\perp c$ geometry.}
\end{figure}

\begin{figure}
\centerline{\includegraphics[angle=0,width=0.45\textwidth]{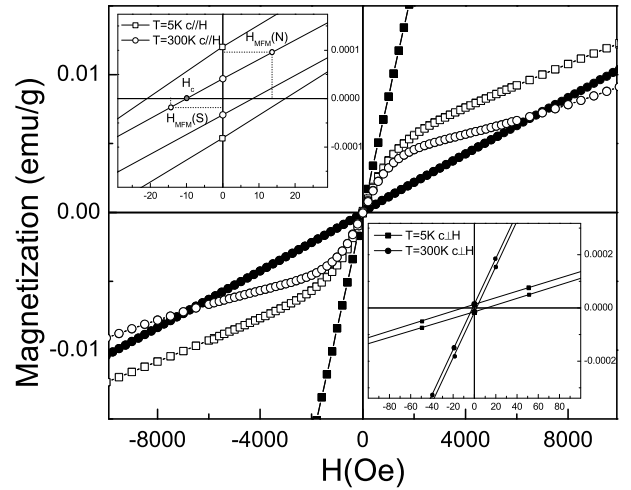}}
\bigskip
\caption{\label{fig02a}Magnetization of the $Zn_{1-x}Co_xO$ film with $x=15\%$ versus magnetic field. $M(H)$ were taken at $5$ (squares) and $300$ K (circles) with $H\perp c$ (filled symbols) and $H\parallel c$ (open symbols) geometry of a magnetic field, respectively. The inserts show in more details the region near the zero field for both geometries.}
\end{figure}

\begin{figure}
\centerline{\includegraphics[angle=0,width=0.45\textwidth]{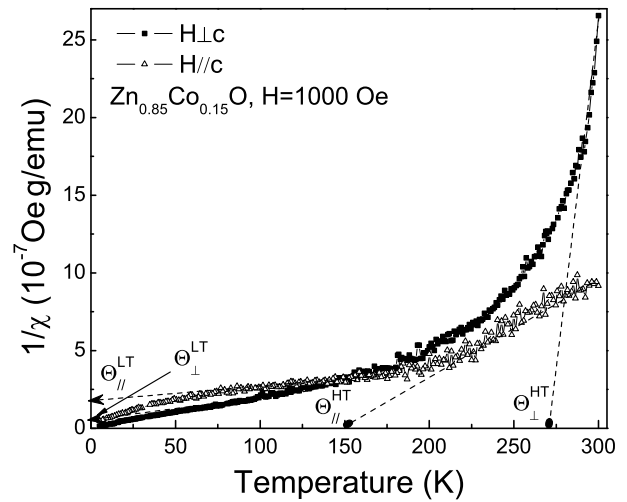}}
\bigskip
\caption{\label{fig03a}Inverse magnetic susceptibility of the $Zn_{1-x}Co_xO$ film with $x=15\%$ versus temperature at a magnetic field of $H=1000$ Oe with $H\perp c$ (filled square) and $H\parallel c$ (open triangle) geometry.}
\end{figure}

In order to study the magnetic properties of the films, we have performed SQUID measurements. 
It is seen from Fig.~\ref{fig02aa} that the temperature dependences for undoped and $15\%Co$-doped $ZnO$ films drastically differ. The observed higher value for magnetization of $15\%Co$-doped $ZnO$ film is caused by strong interaction between $Co^{2+}$ ions with the magnetic moment close to $3\mu_B$ per $Co^{2+}$ ion in $Zn_{1-x}Co_xO$ film~\cite{Sati,Morhain}. The magnetic susceptibility in the undoped $ZnO$ film is negative (Fig.~\ref{fig02aa}) and practically does not depend on temperature. For the $15\%Co$-doped sample one can observe a strong dependence of the magnetic susceptibility on temperature that is described with the Curie-Weiss law. The diamagnetic contribution $\chi=C/(T-\Theta)+\chi_{sub}$, where $C$ is the Curie constant, $\Theta$-Curie-Weiss temperature, and $\chi_{sub}$ - diamagnetic susceptibility of substrate, were taken into account when analyzing SQUID data for the $Co$-doped samples. As a result, for the undoped $ZnO$ film we obtained low magnetization values $<0.1\mu_B$ magnetic moment per defect, which is often related with oxygen vacancies~\cite{Wang1}.  Weak ferromagnetism for undoped $ZnO$ film is clearly demonstrated by magnetization reversal loops in the insert of Fig.~\ref{fig02aa}. For the case of $Co$-doped $ZnO$ films, the obtained results for magnetization are shown in Figs.~\ref{fig02a} and \ref{fig03a} with account of the substrate diamagnetic contribution.

Peculiarities of hysteresis curves of the $Zn_{1-x}Co_xO$ samples are observed in magnetization measurements up to $300$ K (Fig.~\ref{fig02a}). The temperature dependence of inverse magnetic susceptibility in a magnetic field of $1000$ $Oe$ is shown in Fig.~\ref{fig03a}.
The magnetic susceptibility has two well distinguished temperature regimes (at low (LT) and high (HT) temperatures, respectively) with a typical Curie-Weiss behavior for both $H\perp c$ and $H\parallel c$ geometry of magnetic field. The Curie temperatures $\Theta$ obtained from extrapolation to the temperature axis show clearly the HT and LT regimes of effective magnetic interactions (Fig.~\ref{fig03a} and Table \ref{CWeiss}).
As a result of the dominant ferromagnetic interactions between the $Co^{2+}$ ions one can observed hysteresis loops with coercivity values of $H_c^\perp \approx20$ $Oe$ and $H_c^\parallel \approx10$ $Oe$ at $300$ K (inserts on Fig.~\ref{fig02a}). On the other hand, at low temperature ($T<200$ K), one  obtains a negative Curie-Weiss temperatures (Fig.~\ref{fig03a} and Table \ref{CWeiss}), which can be considered as a result of antiferromagnetic behavior of $Co-O-Co$ sequences and has  previously been observed for the $Zn_{1-x}Co_xO$ films~\cite{Morhain,Ney} and powders~\cite{Risbud}.
So, when analyzing the magnetic properties of Co-doped $ZnO$ films, the change in magnetic behavior of the $Zn_{1-x}Co_xO$ films can be understood as the result of a competition between the ferro- and antiferromagnetic $Co$ interactions, which are due to isolated $Co$ ions and $Co-O-Co$ sequences~\cite{Ney, Kobayashi}, respectively. Effective indirect exchange interaction between isolated $Co^{2+}$ ions decreases with decreasing the carrier mobility and at low temperatures the dominant magnetic interaction is antiferromagnetic due to Co-O-Co sequences. The carrier concentration in our films is $\sim10^{20}$ $cm^{-3}$ at any temperature, a value obtained from the $\omega^+$ LOPCM in the Raman spectra. Therefore for these high quality MBE-grown $Zn_{1-x}Co_xO$ films, the high-temperature ferromagnetism is expected to be due to the high electron concentration in the conduction band.

\begin{table}[h]
\caption{\label{CWeiss}Values of the Curie-Weiss temperatures at magnetic field geometries of $H\perp c$  and $H\parallel c$ for of the $Zn_{1-x}Co_xO$ film with $x=15\%$.}
\begin{tabular}{ccccccc}
\hline
\hline
\\
$\Theta^{LT}_\parallel$ (K) & & $\Theta^{LT}_\perp$ (K) & & $\Theta^{HT}_\parallel$ (K) & & $\Theta^{HT}_\perp$ (K) \\
\hline
\\
$-200$ & & $-40$ & & $150$ & & $270$\\
\hline
\hline
\end{tabular}
\end{table}

It is well known that for $Zn_{1-x}Co_xO$ films the magnetization curves at low magnetic field, which is a more favorable for ferromagnetism observation, can be different from the prediction of the effective spin model~\cite{Morhain} used for an interpretation of the para- and antiferromagnetic $Co$ behaviors. For studied films, the magnetization (Fig.~\ref{fig02a}) and magnetic susceptibility (Fig.~\ref{fig03a}) curves reveal the presence of a significant magnetic anisotropy with a magnetic moment $M\perp c$ as observed in Refs.~\cite{Morhain}, \cite{Ney}. The fact that $M \perp c$ is greater than $M \parallel c$ at a low temperature is in good agreement with theoretical predictions that antiferromagnetic interactions of Co-O-Co sequences are less stable than ferromagnetic interactions along the $c$ direction~\cite{Eun-Cheol,Zhang}. The ferromagnetic interaction in this direction is more favorable for ferromagnetism, at least for temperature above $200$ $K$, as observed in our SQUID measurements (Fig.~\ref{fig03a}).

\begin{figure}[h]
\centerline{\includegraphics[angle=0,width=0.45\textwidth]{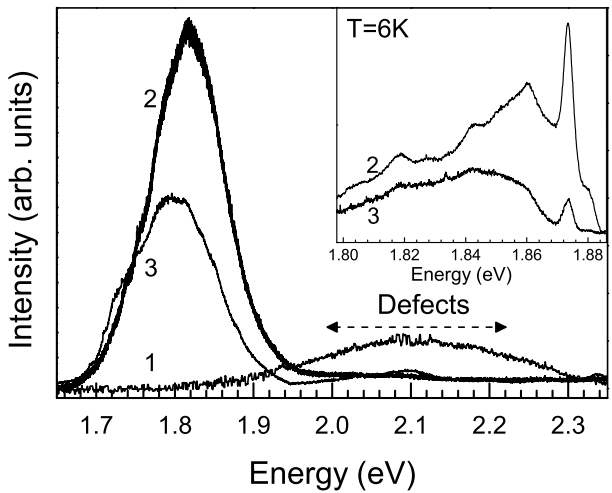}}
\bigskip
\caption{\label{fig02}PL spectra of the undoped (1), $5$ and $15\%Co$-doped (2 and 3, respectively) $ZnO$ films excited by $E_{exc}=2.54$ $eV$ ($488.0$ nm) at  $T=300$ K and  $T=6$ K (insert).}
\end{figure}

Fig. \ref{fig02} provides PL spectra of undoped (curve 1) and $Co$-doped ($5$ and $15\%Co$, curve 2 and 3, respectively) $ZnO$ films at $T=300K$ (in insert at $T=6$ $K$).
In the PL spectra of undoped $ZnO$ films, a broad green emission band is observed at $\sim2.18$ $eV$, which is associated with intrinsic deep-level defects in $ZnO$, namely: oxygen vacancies, interstitial zinc atoms, and antisite oxygen atoms~\cite{Zhang,Coey}.
For the $5\%Co$-doped $ZnO$, the red emission peak at $\sim1.816$ $eV$ ($\sim683$ nm) (Fig.~\ref{fig02}, curve 2) corresponds to electron transitions between  $d$-levels ~\cite{Lommens} of isolated $Co^{2+}$ ions tetrahedrally coordinated to oxygen atoms. Increasing $Co$ concentration to $15\%$, induces a red-emission weakening due to decrease of the isolated $Co^{2+}$ ions and increase of  $Co-O-Co$ sequences~\cite{Chambers,Ney1}, which don't contribute to the red band emission. In addition, the effect of decreasing the intensity of red emission cannot be related with formation of secondary phases such as octahedral $Co$ oxides, since no indication of additional structure phases were observed in Raman and X-ray diffraction measurements within the detection limit.

\begin{figure}
\centerline{\includegraphics[angle=0,width=0.45\textwidth]{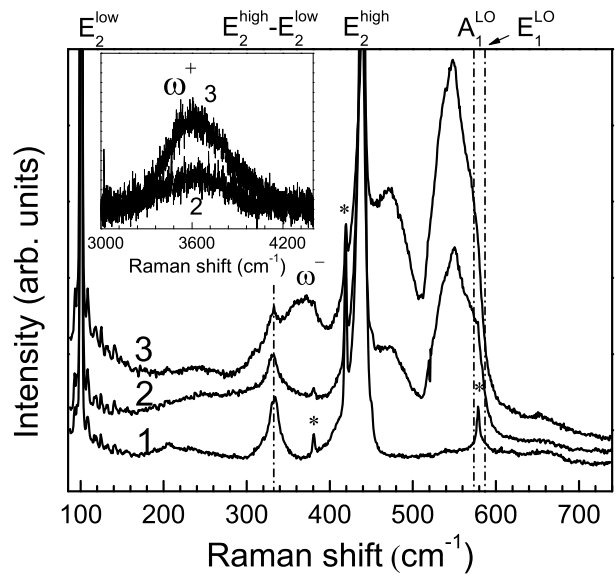}}
\bigskip
\caption{\label{fig03}
Room temperature Raman spectra of undoped (1) and doped with $5$ and $15\%Co$ (2 and 3, respectively) $ZnO$ films. $E_{exc}=2.54$ eV. $T=300$ K.}
\end{figure}

Raman measurements were performed to analyze the vibrational modes and lattice structure of the $Co$-doped $ZnO$ films. They confirm that the films do have the wurtzite structure. Indeed, Fig.~\ref{fig03} exhibits micro-Raman spectra taken from an undoped (curve 1) and $Co$-doped ($5$ and $15\%Co$, curves 2 and 3, respectively) $ZnO$ films. According to group theory, four Raman-active modes $A_1$, $E_1$ and $2E_2$ ($E_2^{low}$ and $E_2^{high}$) are expected for the wurtzite-type $ZnO$ structure, which belongs to the space group $P6_{3}mc$. The polar nature of $A_1$ and $E_1$ modes leads to a splitting into $TO$ and $LO$ components. The $E_2^{low}$ and $E_2^{high}$ modes are non-polar. In the backscattering geometry for ($0001$) $ZnO$, both $E_2$ and $A_1(LO)$ modes can be detected. The $A_1(LO)$ mode at $\thicksim574$ $cm^{-1}$ shows very low intensity for high-quality $ZnO$ films. The most pronounced peaks in $ZnO$ originate from $E_2^{low}$ and $E_2^{high}$ phonon modes at $\thicksim100$ and $\thicksim437$ $cm^{-1}$, respectively. The $E_2^{high}$-$E_2^{low}$ modes are observed at $\thicksim333$ $cm^{-1}$. Fig.~\ref{fig03} illustrates also the presence of three phonon modes of the sapphire substrate (denoted asterisks) with the $A_{1g}$ ($418$ $cm^{-1}$) and $E_g$ ($379$ and $578$ $cm^{-1}$) symmetries, respectively.

The Raman non-polar $E_2^{low}$ and $E_2^{high}$ modes in undoped $ZnO$ films are very sensitive to disorder in zinc and oxygen sublattices, respectively. According to Fig.~\ref{fig03} (curve 1), for undoped $ZnO$ films the $E_2^{low}$ mode at $\sim100.5$ $cm^{-1}$, involving mainly $Zn$ motion, displays a very narrow linewidth ($\sim1.6$ $cm^{-1}$). After $Co$ doping, the $E_2^{low}$ mode intensity strongly decreases. The mode is broadened up to $\sim2.4$ $cm^{-1}$ and undergoes a red shift (up to $\sim0.7$ $cm^{-1}$) with respect to undoped $ZnO$. It is the effect of compositional fluctuations induced by random substitution of $Co$ ions into $Zn$ sites in host lattice. Such an “alloying effect” does not usually involve any precipitation of other crystalline phases and occurs, for example, in $(Ga,In)N$, where $In$-rich quantum-dot-like regions arise~\cite{Davydov}. The spectra also exhibit an intense $E_2^{high}$ mode associated with oxygen-atom vibrations which appears at $\sim439.5$ $cm^{-1}$  with a full-width-at half maximum $\Gamma\sim5$ $cm^{-1}$ for undoped $ZnO$ film. With the increase of $Co$ amount, the $E_2^{high}$ mode shows a red shift up to $\sim1.2$ $cm^{-1}$ and broadens up to $\sim13$ $cm^{-1}$ due to disorder effects in the oxygen sublattice (vacancies, interstitials) inducing a change in coordination numbers of some $Co$ atoms due to oxygen vacancies.

In the Raman spectra of most heavily doped $ZnO$ is often observed intensity signal in region between of the $TO$ and $LO$ modes, the interpretation of this bands is ambiguous. For $Co$-doped samples, an additional overlapping broad and intensity bands in the region of $450$-$580$ $cm^{-1}$ are detected (Fig.~\ref{fig03}). These bands were seen in different scattering geometries and therefore could be attributed to the $ZnO$ phonon states due to disorder activated Raman scattering~\cite{Serrano}.
It is assumed also \cite{Chassaing,Fonoberov} that for $Co$-doped $ZnO$ nanostructures a broad feature at $470$-$500$ $cm^{-1}$ may be assigned to the surface optical phonon mode (SOP) (Fig.Fig.~\ref{fig03}). When the crystallite size of $ZnO$ is $10 nm<L<100 nm$, the SOP can be appear and its intensity increasing with reducing the nano-column diameter~\cite{Hayashi}. It is noteworthy that the SOP peak is reliably detected in resonant multi-phonon Raman spectra ($E_{exc}=3.81$) of undoped and $Co$-doped $ZnO$ films, the frequency of this mode being independent of the Co concentration (is not shown). Difficulties in SOP detection in non-resonant Raman spectra of undoped $ZnO$ films may be due to as well with large crystalline size ($150$-$200$ nm) as considerable disordering at the boundaries of the internal grain structure, as compared with $Co$-doped films.

Additional bands at $\sim488$, $550$ and $\sim708$ $cm^{-1}$ in $Co$-doped bulk samples and thin $ZnO$ films grown by various methods have been reported in the literature~\cite{Samanta,Liu1}. These bands arise from secondary structural phases. They can be clusters of $Co_3O_4$ or isometric compounds $Zn_xCo_{3-x}O_4$. However, in the present samples, additional Raman bands of these secondary phases are not present.

An intense wide band appears in the frequency range of $550$-$600$ $cm^{-1}$ for the $Co$-doped samples (Fig.~\ref{fig03}). At least two Lorentzian profiles are necessary for fitting the band which splits into two subbands at frequencies $\sim550.8$ $cm^{-1}$ ($\Gamma\sim 33$ $cm^{-1}$) and $\sim576.3$ $cm^{-1}$ ($\Gamma\sim 22$ $cm^{-1}$) for the $5\%Co$ and $\sim546.2$ $cm^{-1}$ ($\Gamma\sim46$ $cm^{-1}$) and $\sim572.9$ $cm^{-1}$ ($\Gamma\sim27$ $cm^{-1}$) for the $15\%Co$ concentrations, respectively. It is interesting to note that with increasing  $Co$ concentration from $5$ to $15\%$, the intensity of both modes is substantially increased. This gives a clear evidence for the $Co$ substitution in $ZnO$ host lattice~\cite{Samanta}. A similar increase in the intensity of the Raman band was reported for the multiphonon mode at $540$ $cm^{-1}$ and $E_1(LO)$ mode at $584$ $cm^{-1}$ in $Co$-doped~\cite{Samanta} $ZnO$. In our opinion, these bands are related with the resonant Raman effect at subband-gap excitation caused by $d$-$d$ transitions in $Co^{2+}$ ions as well as by defect levels in $ZnO$ host in  the energy range $2.2-3.0$ $eV$~\cite{Hasuike}. The extrinsic Fr\"{o}hlich interaction mediated by the localized electronic states bounded to defect, impurities and $3d$-related levels of $Co^{2+}$ ions could enhance the scattering efficiency independently on the phonon wave vector $\bf{q}$.

The zone-center LO phonons are affected by the $n$-type conductivity which is due to the oxygen vacancy ($V_o$) and interstitial $Zn$($Zn_i$)~\cite{TCZO} of the $Co$-doped $ZnO$ films, since we deal with electron concentrations higher than $10^{19}$ $cm^{-3}$. In polar semiconductors, when the frequency of longitudinal plasma oscillations approaches the $LO$ phonon frequencies, their macroscopic electric fields strongly interact, which results in appearance of the $\omega^-$ and $\omega^+$ LOPCMs. However, owing to poor carrier mobility of the doped $ZnO$ epilayers, it is expected that the LOPCMs are overdamped due to existence of many structural defects.

\begin{figure}
\centerline{\includegraphics[angle=0,width=0.45\textwidth]{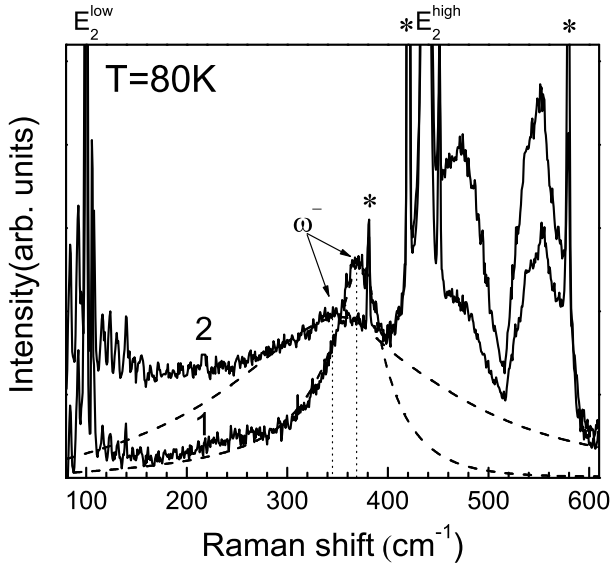}}
\bigskip
\caption{\label{fig04}Raman spectra for the $Zn_{1-x}Co_xO$ films at $80K$. The dashed lines correspond to the modelled $\omega^-$ LOPCMs band at $\sim342$ and $\sim368$ $cm^{-1}$ with $5$ ($2$) and $15\%Co$ ($1$), respectively.}
\end{figure}

In order to assign the bands in  films with $5$ and $15\%Co$ at $\sim342$ $cm^{-1}$ and $\sim368$ $cm^{-1}$, respectively, to plasmon modes, we performed the Raman measurements at temperatures from $80$ to $500$ K (Fig.~\ref{fig04}) analyzed  band shapes with the help of a semiclassical theory of Raman scattering~\cite{Irmer}. Both the electro-optic and deformation potentials ($I^{DP-EO}$) as well as charge-density ($I^{CDF}$) contributions to the processes of light scattering were taken into account. The $\omega^-$ plasma-like modes are fitted by using the following set of equations:
\begin{equation}
\begin{array}{l}\label{EqA}
I(\omega)=A_1I^{DP-EO}(\omega)+A_2I^{CDF}(\omega)=\\
\\
\left(A_1\overline{\left[\frac{\omega^2_{TO}(1+C_{FH})-\omega^2}{\omega^2_{TO}-\omega^2}\right]^2}+A_2\overline{\left[\frac{\omega^2_{LO}-\omega^2}{\omega^2_{TO}-\omega^2}\right]^2}\right)Im\left(-\frac{1}{\epsilon(\omega)}\right),\\
\\
\epsilon(\omega,\vec{q})=\epsilon_\infty+\epsilon_\infty\frac{\omega^2_{LO}-\omega^2_{TO}}{\omega^2_{TO}-\omega^2-i\Gamma\omega}-\epsilon_\infty\frac{\omega^2_p}{\omega(\omega+i\gamma)},
\end{array}
\end{equation}
where $\epsilon_\infty$ is the high-frequency dielectric constant, $C_{FH}$ is the Faust-Henry coefficient, $\omega_{LO}$ and $\omega_{TO}$ are frequencies of the $LO$ and $TO$ phonons, $\Gamma$($\gamma$) is a phonon (plasmon) damping coefficient, $\omega_p$ is a plasmon frequency.

We used the prefactors  of $Im(-1/\epsilon)$ in Eq.~\ref{EqA} for the light-scattering efficiency from Ref.~\cite{Irmer}. By using the optimization procedure for the $\omega^-$ plasma-like mode~\cite{Bryksa} one finds the $A_1, A_2$ coefficients and the $\gamma$ parameter for which the sum of the chi-square value, $\sum(I^{exp}(\omega)-I(\omega))^2$ is  minimal at fixed $\omega_p$ and $\Gamma$ values. The $\omega^-$ band shape fitting analysis was made for each of the Raman spectra at a given temperature in order to get the plasmon damping, $\gamma$ versus temperature.

Two calculated LOPCM’s modes (dashed lines) are shown in Fig.~\ref{fig04} and demonstrate very good agreement with the experimental spectra. For modeling the $\omega^-$ LOPCM band we used the following parameter values: $C=6.4$~\cite{Bairamov}, $\omega_{TO}=381.6$ $cm^{-1}$ and $\omega_{LO}=574.2$ $cm^{-1}$, $\epsilon_\infty=3.67$ and the effective mass of electron $m^*=0.27m_0$ where m0 is the electron mass in vacuum. The plasmon damping value provides the carrier mobility value ($\mu=e/m^*\gamma$), and the plasma frequency $\omega_p$ is related with the carrier concentration $n$ by the relation $\omega_p^2=4\pi e^2n/\epsilon_\infty m^*$. Therefore, one can obtain also the carrier mobility value, $\mu$, versus temperature.

\begin{figure}
\centerline{\includegraphics[angle=0,width=0.45\textwidth]{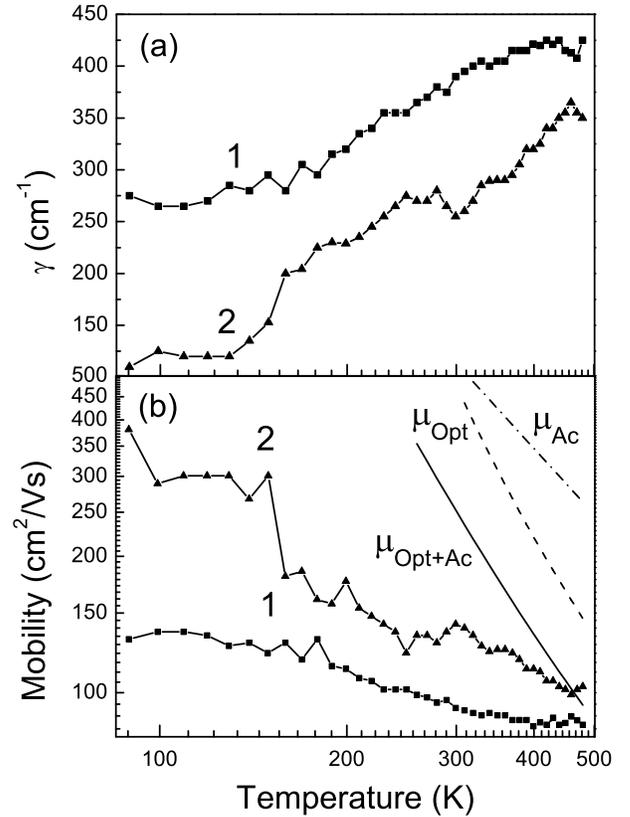}}
\bigskip
\caption{\label{fig05} Temperature dependence of the plasmon damping(a) and electron mobility(b) obtained from the analysis of Raman spectra for $Zn_{1-x}Co_xO$ films with $5\%$ ($1$) and $15\%$ ($2$). The solid, dashed and dot dashed lines show the dependences for the electron mobilities limited by processes of scattering on acoustic and optic phonons as well as limited by their joint contribution, respectively. $\omega_p=3400$ $cm^{-1}$. $\Gamma=47$ $cm^{-1}$.}
\end{figure}

Fig.~\ref{fig05} shows the temperature dependence of the electron mobility for the $Zn_{1-x}Co_xO$ films with $5$ and $15\%Co$, respectively, obtained from modeling the $\omega^-$ LOPCM’s band. We found $\omega_p$ and $\Gamma$ values equal to $3400$ $cm^{-1}$ and $47$ $cm^{-1}$, respectively, at any temperature. The value of $\omega_p$ corresponds to electron concentration $\sim1.3 \times 10^{20}$ $cm^{-3}$. Such large value for $\omega_p$ is in  good agreement with a spectral position of the $\omega^+$ LOPCMs ($\omega_p\approx\omega^+$) observed in the experimental Raman spectra for the $Zn_{1-x}Co_xO$ films with $5$ and $15\%Co$ (see insert in Fig.~\ref{fig03}). The plasmon damping parameter $\gamma$ has a strong temperature dependence, which arises from the temperature dependence of the electron mobility. In order to determine the influence of ferromagnetic ordering on the carrier mobility, we calculate contributions to the mobility, which are due to the carrier scattering process on the acoustic $\propto(kT)^{-3/2}$ and optic $\propto(exp(\hslash\omega_{LO}/kT)-1)$ phonons (Fig.~\ref{fig05}) in high quality epitaxial undoped $ZnO$ films~\cite{TCZO}. Even if the mobility in our $Zn_{1-x}Co_xO$ films decreases at temperature increasing up to $500$ K,  its value is comparable to that in structurally perfect $ZnO$ films at temperatures around 500 K.
It is interesting to note that the electron mobility in the films with $15\%Co$ is higher than that with $5\%Co$ at any temperature  (Fig.~\ref{fig05}). Correlation between magnetic and transport properties was published for DMS based on $A^3B^5$ semiconductors~\cite{Jungwirth}. For example, for p-$GaMnAs$ the maximum value of Curie temperature ($T_c=110$ K) was obtained for a metallic type conductivity and with a higher value of the charge carrier mobility~\cite{Jungwirth}. Such correlation between magnetic properties and electron mobility takes place in the studied $Zn_{1-x}Co_xO$ films, too (Fig.~\ref{fig05}).

Up to now the physical mechanism of ferromagnetic ordering in $n$-type $Zn_{1-x}Co_xO$ is not ascertained yet. One can offer the following microscopic mechanism of ferromagnetism for the electron concentration put into evidence in the conduction band. The mechanism foresees long-range interaction between two localized magnetic moments $\vec{S_i}$ and $\vec{S_j}$ of isolated $Co^{2+}$($S=3/2$) ions at a distance of $\vert\vec{R_i}-\vec{R_j}\vert$ via free electrons in the conduction band. For these magnetic moments, an important role is plaid by the parameter of exchange interaction $J_{ij}$ that can oscillate in the direct space, as for $RKKY$ interaction mechanism~\cite{Jungwirth,Bryksa1}. The exchange interaction $J(\vert\vec{R_i}-\vec{R_j}\vert)$, which is responsible for the Curie temperature, depends on the electronic subsystem of $Zn_{1-x}Co_xO$ semiconductor~\cite{Bryksa1}.
For the samples to be ferromagnetic, most of the $Co$-atom spins should be parallel one to another. In other words, the symmetry of the total system including the crystalline lattice and Co-atom spins should be higher in the ferromagnetic phase. 

Since the symmetry of the total system for the ferromagnetic phase is maximal, one should expect larger electron mobility than for a disordered configuration. However, the situation is more complex for studied anisotropic wurtzite $Co$-doped $ZnO$ films. There takes place considerable anisotropy of magnetization with the easy-axis magnetization $H\perp c$~\cite{Sati,Morhain} (Fig.~\ref{fig02a}). For ferromagnetic phase of $Co$-doped $ZnO$ films ($H\perp c$, $T>40K$, Table~\ref{CWeiss}), we performed the analysis of Raman spectra of LOCPMs and found an increase of the carrier mobility at decreasing temperature (Fig.~\ref{fig05}). This effect could be explained by phonon mechanism of the carrier scattering~\cite{TCZO}. Note that the carrier mobility (and magnetization) is higher for $15\%Со$ sample as compared with $5\%Со$ sample (Fig.~\ref{fig05}). It remains to explain why ferromagnetism takes place only at rather high temperatures. One can suggest that coupling between electron spins and $Co$-atom spins is favored by collisions with phonons.  Phonon population increases with temperature which increases collision probability.

\section{Summary and Outlooks}

In this work, we have studied magnetic, structural, optical and electronic properties of high quality MBE-grown $Zn_{1-x}Co_xO$ films with $x=0$ (undoped), $x=5\%$ and $x=15\%$. We provide  experimental evidence for important role of electrons, which rise ferromagnetism properties up to room temperature.

From MFM magnetization maps and SQUID measurements, ferromagnetic behavior of films at room temperature is clearly put into evidence. SQUID data show a complicated temperature dependence of the magnetic susceptibility, which is due to two different kinds of coordination, at the range of second first neighbors, for $Co^{2+}$ ions within the $ZnO$ host. $Co-O-Co$ sequences contribute to antiferromagnetic behavior whereas isolated $Co^{2+}$ ions contributes to ferromagnetic properties of the films. High temperature ferromagnetism results from interaction between isolated ions at cation sites mediated by conduction electrons (RKKY-like mechanism). On the contrary, at low temperatures (temperatures below $150K$), the antiferromagnetic effect of $Co-O-Co$ sequences is dominant.

The Raman measurements confirm the high crystalline quality of both undoped and $Co$-doped $ZnO$ films as well as their wurtzite structure. Raman bands of the antiferromagnetic $Co$ oxygen spinel clusters have not been observed. The red $Co$ emission exhibit a broad peak at $1.816$ eV ($683$ nm), which can be ascribed to electron transitions within isolated $Co^{2+}$ ion.

Raman investigation of LOPCMs versus temperature has been used to probe the free-carrier properties in films. A modeling of the $\omega^-$ LOPCM’s band was performed which allows determining the temperature dependence of the charge carrier mobility. Curie temperature increases with $Co$ concentration from $5\%$ to $15\%$, and the magnetic films with a higher value of the magnetisation have a higher electron mobility.
\\

\section{Acknowledgments}

This work has been performed within Grant 21344 FW from Minist\`{e}re des Affaires Etrang\`{e}res (France).

\end{document}